\documentclass[aps,prb,reprint,longbibliography,superscriptaddress,floatfix]{revtex4-2}
\usepackage[utf8]{inputenc}

\usepackage{csquotes}
\usepackage[british]{babel}
\usepackage{amssymb,amsmath}
\usepackage{graphicx}
\usepackage{hyperref}
\hypersetup{
  final,
  pdfborder={0 0 0},
  pdftitle={Atomic resolution observations of silver segregation in a [111] tilt grain boundary in copper},
  pdfauthor={Langenohl, Brink, Richter, Dehm, Liebscher},
  colorlinks=true,
  urlcolor=blue,
  linkcolor=blue,
  citecolor=blue
}
\usepackage{booktabs}
\usepackage[dvipsnames]{xcolor}
\usepackage[separate-uncertainty=true]{siunitx}
\usepackage{float}
\usepackage[normalem]{ulem}

\begin{document}
\frenchspacing

\title{Atomic resolution observations of silver segregation in a [111] tilt grain boundary in copper}

\author{Lena Langenohl}
\thanks{L.L. and T.B. contributed equally to this work.}
\affiliation{Max-Planck-Institut f\"ur Eisenforschung GmbH, Max-Planck-Stra\ss{}e 1, 40237 D\"usseldorf, Germany}

\author{Tobias Brink}
\thanks{L.L. and T.B. contributed equally to this work.}
\email{t.brink@mpie.de}
\affiliation{Max-Planck-Institut f\"ur Eisenforschung GmbH, Max-Planck-Stra\ss{}e 1, 40237 D\"usseldorf, Germany}

\author{Gunther Richter}
\affiliation{Max Plank Institute for Intelligent Systems, Heisenbergstr. 3, 70569 Stuttgart, Germany}

\author{Gerhard Dehm}
\email{dehm@mpie.de}
\affiliation{Max-Planck-Institut f\"ur Eisenforschung GmbH, Max-Planck-Stra\ss{}e 1, 40237 D\"usseldorf, Germany}

\author{Christian H. Liebscher}
\email{liebscher@mpie.de}
\affiliation{Max-Planck-Institut f\"ur Eisenforschung GmbH, Max-Planck-Stra\ss{}e 1, 40237 D\"usseldorf, Germany}

\date{\today}

\begin{abstract}
Alloying a material and hence segregating solutes to grain boundaries is one way to tailor a material to the demands of its application. Direct observation of solute segregation is necessary to understand how the interfacial properties are altered. In this study, we investigate the atomic structure of a high angle grain boundary both in pure copper and upon silver segregation by aberration-corrected scanning transmission electron microscopy and spectroscopy. We further correlate the experiments to atomistic simulations to quantify the local solute excess and its impact on grain boundary properties. We observe that the grain boundary structure remains intact upon silver segregation and up to five different positions within a structural unit serve as segregation sites. By combining the atomic resolution observation with atomistic modelling, we are able to quantify the local silver concentration and elucidate the underlying segregation mechanism.
\end{abstract}

\maketitle

\newcounter{supplfigctr}
\renewcommand{\thesupplfigctr}{\arabic{supplfigctr}}
{\refstepcounter{supplfigctr}\label{suppl_fig:EBSD-comparison-before_after_annealing}}
{\refstepcounter{supplfigctr}\label{fig:DSC-lattice}}
{\refstepcounter{supplfigctr}\label{fig:Asymmetric_GB}}
{\refstepcounter{supplfigctr}\label{fig:Ag_occupancy_500K}}
{\refstepcounter{supplfigctr}\label{fig:Eseg-atom-vs-col}}
{\refstepcounter{supplfigctr}\label{fig:sideview}}
{\refstepcounter{supplfigctr}\label{fig:steinhardt}}

\section{Introduction}
Segregation of elements to grain boundaries (GBs) has a dramatic impact on material properties. Examples include the material becoming stronger, or generally having a more stable microstructure ~\cite{Ozerinc2012,Koju2020,Basu2016,Khang2006,Schaefer2012,Rupert2011,Vo2011}, but also increasing the susceptibility to failure through corrosion or embrittlement~\cite{Wei1986,Seah1980,Lejcek2010}. One of the most prominent examples of embrittlement is alloying bismuth to copper, which was already discovered back in 1874~\cite{Hampe1874}. %
More recent publications explain the embrittlement by a weakening of the metallic bonds at the GB, which can be related to a pure atomic size effect or the change of the electronic structure at the GB~\cite{Duscher2004,Keast1998, Schweinfest2004}.

Early research on segregation revealed significant changes in segregation behaviors for GBs with different misorientations between both grains using autoradiographs, atomic absorption spectroscopy or Auger electron spectroscopy~\cite{Ainslie1960,Thomas1955,Li1995,Powell1973}. Even though such techniques can reveal chemical changes by less than 1 at-$\%$ and the spatial resolution across the GB can go down to \SI{0.1}{nm}, the lateral resolution, revealing chemical changes within the GB structure, is impossible. Furthermore, it was already stated since a long time that also other GB parameters and thus the existing GB and its atomic structure should play a crucial role, even though experimental observations on the atomic scale were not available~\cite{Hondros1975, Watanbe1985,Vitek1982,Wynblatt2006,Lejcek2018}. Segregation can change different properties of the existing GB phase, such as its mobility, energy, stability, but also its atomic structure~\cite{Raabe2014}. It can introduce mono- or bilayers~\cite{Cantwell2020,Duscher2004,Frolov2015} or precipitates of solute atoms at the GB, and can even change the GB plane by faceting or induce a GB phase transformation ~\cite{Ference1988,Meiners2020,Peter2018,Peter2021}. Only few direct experimental observations of segregation at the GB on the atomic level exist so far since high-resolution imaging microscopes and spectroscopy are needed~\cite{Khang2006,Peter2018,Peter2021,Feng2016,Ma2013,Yoon2017,Li2014,Parajuli2019,Duscher2004,Nie2013}. To completely comprehend how segregation affects macroscopic features, such as strengthening, embrittlement, or corrosion, we must first identify the segregation-induced atomistic changes inside a GB structure.

The fundamental thermodynamics implemented by Gibbs can be used to describe adsorption at interfaces. Langmuir and McLean used these equations to develop a model of GB segregation, describing the concentration limit of solutes at the GB in dependence of the bulk concentration and temperature~\cite{McLean1957}. 
Their model combines thermodynamics with statistical mechanics, using a finite number of atomic positions in the GB which can be completely replaced by solute atoms~\cite{McLean1957}.  %
Since it is based on mean field theory, only a single value of GB segregation concentration can be calculated and it does not take into account site-specific changes of concentrations within the GB structure. It neglects interactions between individual atoms, which can significantly contribute to the segregation behavior~\cite{Hondros1977}. Guggenheim and Fowler extended the segregation model, taking into account the interaction between neighboring atoms and the probabilities of neighboring atoms being a solute or matrix element~\cite{Fowler1939}. Recently, more and more attention was paid to site-specific segregation within the GB structure~\cite{White1977,White1978,Udler1998,Wagih2020,Huber2018}. In order to proof thermodynamic models which take that into account, experimental observations are needed, which can validate the theoretically calculated GB structure itself and the segregational site occupancies~\cite{Lejcek2018}.

Due to its high conductivity, Cu is used in many electronic applications and long-pulse high field magnets~\cite{Debray2013}. But because Cu is also quite soft and would not sustain the electromagnetic forces, alloying different elements is a common technique to increase its tensile strength which needs to be balanced with an undesired concomitant decrease in conductivity~\cite{Zherlitsyn2012}.  When alloying Ag to Cu, an increase in tensile strength can be observed~\cite{Vo2011,Lozovoi2006}. Upon specific heat treatments, it can be accompanied with a high conductivity of $\geq$80$\%$ IACS (International Annealed Copper Standard)~\cite{Sakai1991,Sakai1997,Yang2020}. 

Even though Cu-Ag alloys are promising materials, only few experimental studies reveal the segregation of elements at the GB on an atomistic level~\cite{Peter2018,Peter2021}. Divinski et al. investigated the diffusion behavior of Ag in Cu $\Sigma$ 5 GBs and observed abrupt changes in diffusivity at elevated temperatures, attributed to a GB phase transformation~\cite{Divinski2012,Frolov2015}.  Simulation studies on Cu-Ag systems can explain underlying atomistic effects~\cite{Frolov2013,Lozovoi2006,Vo2011,Koju2020,Koju2020CuAg,Frolov2015}. However, experimental observations are needed to confirm the atomistic changes and structures at the GB. Cu-Ag is a good model system to study segregation at GBs due to its extended miscibility gap and absence of intermetallic phases~\cite{Subramanian1993}.

Here, we investigate Ag segregation at a $\Sigma$37c $\langle111\rangle \{347\}$ GB at atomic resolution using aberration corrected scanning transmission electron microscopy (STEM) combined with atomically resolved energy dispersive X-ray spectroscopy (EDS) and hybrid molecular dynamics (MD)/Monte Carlo (MC) simulations. We show different methods for correlating experimental observations with simulations and investigate the evolution of Ag segregation as a function of its concentration. Finally, we discuss different parameters which may be responsible for the site occupation of Ag within a high angle GB.

\section{Methods}

\subsection{Experimental specimen preparation and STEM investigation}

A Cu thin film was deposited on a $\langle0 0 0 1\rangle$ sapphire wafer by the Central Scientific Facility Materials of the Max Planck Institute for Intelligent Systems in Stuttgart. The Al$_2$O$_3$ substrate was sputter cleaned at \SI{200}{eV} for \SI{5}{min} and subsequently annealed at \ang{1000}C for 1h. In a next step, a \SI{2}{\micro\meter} thick layer of Cu was deposited at room temperature by molecular beam epitaxy (MBE) with a rate of \SI{0.3}{nm/s}. After deposition, the film was annealed for \SI{2}{h} at \SI{450}{\celsius} inside the same chamber.
For the investigation of Ag segregation, \SI{100}{nm} of Ag were sputtered on top of the film at the Ruhr-Universit\"at Bochum. Prior to that, the top oxide layer was removed by Ar$^+$ sputtering. Consecutively, the film covered with Ag was annealed at \SI{600}{\celsius} for \SI{16}{h}. For comparison, one part of the same film was annealed at \SI{600}{\celsius} for \SI{16}{h} without a Ag layer.
Both films, with and without Ag, were characterised using a Thermo Fisher Scientific Scios2HiVac dual-beam secondary electron microscope (SEM) equipped with an electron-backscattered diffraction (EBSD) detector. Inverse pole figure (IPF) maps of the EBSD scans helped to identify the GBs of interest, which were then lifted out using a focused Ga$^+$ ion beam (FIB). The rough milling was performed at \SI{30}{kV} and \SI{5}{nA}, going sequentially down to \SI{5}{kV} and \SI{7.7}{pA} up to a thickness of about \SI{40}{nm} to \SI{80}{nm}.

The FIB lamellas were then inspected using a probe-corrected FEI Titan Themis 80-300 (Thermo Fisher Scientific) (scanning) transmission electron microscope ((S)TEM). The TEM is equipped with a  high-brightness field emission gun, set to a voltage of \SI{300}{kV} and a probe current of 70--\SI{80}{pA} for imaging. A high-angle annular dark-field (HAADF) detector (Fishione Instruments Model 3000) was used to record the HAADF-STEM datasets with collection angles of 78--\SI{200}{mrad} and a semi-convergence angle of \SI{17}{mrad}. 
EDS measurements were performed at \SI{120}{kV} and acquired for at least \SI{15}{min}. As we registered a \SI{120}{kV} STEM-HAADF image along with the EDS measurement and wanted to ensure atomic resolution, a fine scan step size of 512x172 px with a pixel size of \SI{14.47}{pm} was chosen. For the EDS intensity, subsequent binning of 4x4 pixels was needed to obtain sufficient data for each position. An additional gaussian filter was used to make intensity changes in the Ag L-$\alpha$ peak more visible, ensuring at the same time that the main findings stayed unaffected. 

\begin{figure*}
    \centering
     \includegraphics[width=\linewidth]{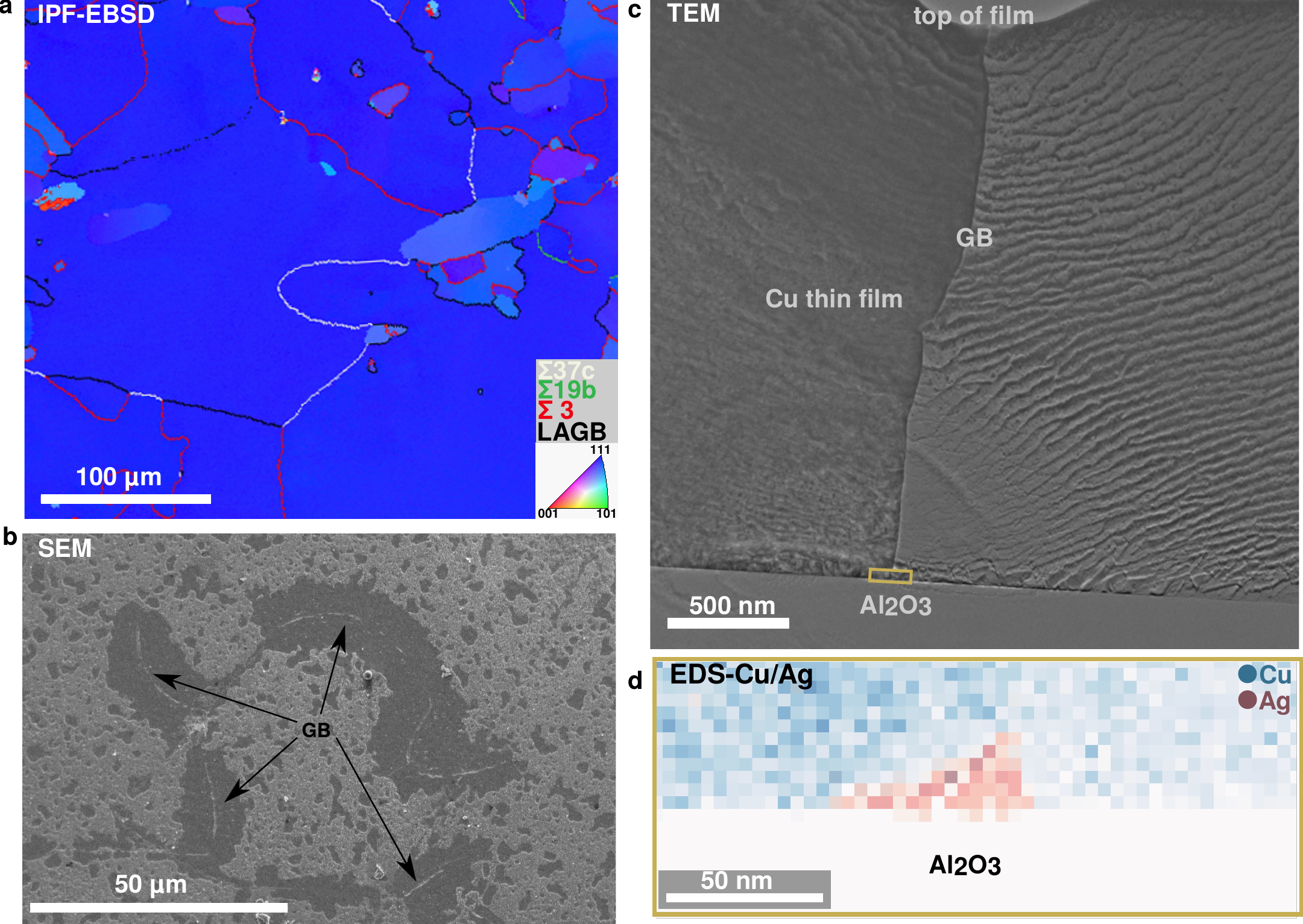}
    \caption{Characterization of the film with and without Ag. a) EBSD of the microstructure of the pure Cu film. b) SEM image of the surface of the film after Ag deposition and annealing. c) Bright field-TEM image of a cross section of a $\Sigma$ 37c GB of the Cu thin film with Ag. d) EDS map from the interface of the GB with the substrate shows a Ag precipitate at the bottom, which signifies that Ag diffused entirely across the GB down to the substrate.}
    \label{fig:ebsd-microstructure}
\end{figure*}

\subsection{MD/MC simulations}

We used LAMMPS \cite{Plimpton1995,Thompson2022} with an embedded atom method (EAM) potential for Cu--Ag by Williams et al. \cite{Williams2006} for all simulations. In a first step, a bicrystal containing the experimentally-observed GB was obtained by joining two copper fcc crystals to produce a $\Sigma37$c GB with $[\overline{1}11]$ tilt axis and $(\overline{3}\overline{7}4)$ and $(3\overline{4}7)$ GB planes, respectively. We varied the relative displacements between the crystallites followed by minimization of the potential energy until the experimentally-observed structure was found ($\gamma$-surface method). The final system was approximately of size $94\times81\times\SI{187}{\angstrom^3}$ with 119\,880 atoms. We applied open boundaries normal to the GB and otherwise periodic boundary conditions.

In a second step, we computed the bulk silver concentration. Alloying was simulated in an isobaric semi-grand-canonical ensemble by hybrid MD/MC simulations \cite{Sadigh2012}, defined by the total number of atoms $N$, pressure $p = \SI{0}{Pa}$, temperature $T$, and chemical potential difference $\Delta \mu = \mu_\text{Ag} - \mu_\text{Cu}$. The latter controls the concentration of silver. MD time integration with a time step length of \SI{2}{fs} was alternated with $0.1N$ MC trial steps every 20 MD time steps. The bulk silver concentration was computed by running the simulations with $\Delta \mu$ from \SI{0.3}{eV} to \SI{1.0}{eV} at \SI{300}{K} and \SI{500}{K} on a fully periodic system of 32\,000 atoms ($20\times20\times20$ unit cells) of fcc copper for \SI{10}{ns}.

Finally, we simulated the segregation of silver to the GBs. In order to avoid size effects due to the excess GB stress \cite{Frolov2012a}, we fixed the sizes of the systems containing the GBs in the periodic directions using the lattice constant of the pure copper system at the given temperature, thus switching to the standard isochoric semi-grand-canonical ensemble. This is a reasonable approximation due to the very low solubility of Ag in Cu, meaning that the lattice constant is not a function of $\Delta \mu$. The lattice constants were obtained by equilibrating a periodic, pure fcc Cu system in the isothermal--isobaric ensemble for \SI{250}{ps}. Then, the system containing the GB was scaled to the correct lattice constant and equilibrated in the semi-grand-canonical ensemble for \SI{20}{ns} with the same parameters as above. At \SI{300}{K} we used $\Delta\mu$ values from \SI{0.30}{eV} (almost pure copper) to \SI{0.74}{eV} (phase transition to silver at \SI{0.72}{eV}). At \SI{500}{K} we used $\Delta\mu$ values from \SI{0.40}{eV} (very low silver concentration) to \SI{0.68}{eV} (GB starts to disorder). Excess properties were calculated as defined in Refs.~\cite{Frolov2012, Frolov2012a} and simulation results were visualized with OVITO \cite{Stukowski2010}. The Ag occupation of atomic columns as viewed from the tilt axis direction was calculated by averaging over a MD/MC simulation running for an MD time of \SI{1}{ns}.

\section{Results and Discussion}

\subsection{Microstructure of the investigated thin film}

We studied the global thin film microstructure and grain boundary evolution by EBSD and TEM combined with EDS before and after Ag segregation. An inverse pole figure map of the electron backscattered diffraction dataset of the pure Cu thin film is shown in Fig.~\ref{fig:ebsd-microstructure}a. The out of plane orientation of most grains is close to $\langle111\rangle$ as expected for Cu on a single crystalline $\langle0001\rangle$ Al$_2$O$_3$ wafer~\cite{Katz1968,Dehm2005}. The Cu grains are oriented in orientation relationship II, in which the $\{111\} \langle0\bar{1}1\rangle$ direction of the Cu film is oriented parallel to the $ (0001) \langle2\bar{1}\bar{1}0\rangle$ direction of $\alpha$-Al$_2$O$_3$ substrate~\cite{Dehm2005}. The grain size varies between several \si{\micro\meter} up to more than \SI{200}{\micro\meter}, all of them being larger than the film thickness (\SI{2}{\micro\meter}). The thin film was then divided into several pieces. One piece was annealed at \SI{600}{\celsius} for \SI{16}{h} and no further grain growth could be observed (see Suppl. Fig.~\ref*{suppl_fig:EBSD-comparison-before_after_annealing} a and b). On another piece we deposited a \SI{100}{nm} layer of Ag. Subsequent annealing under the same condition as for the piece without Ag deposition let Ag diffuse along the GBs. 
An SEM image of the film after Ag deposition and annealing is shown in Fig.~\ref{fig:ebsd-microstructure}b. Ag is still present in most areas of the film (here visible in brighter colors), whereas some grooved GBs show a Ag-depleted zone around the GB (indicated by arrows). To verify that Ag diffused along the complete GB, a cross section of a $\Sigma$ 37c GB was investigated by TEM (Fig.~\ref{fig:ebsd-microstructure}c) and EDS (Fig.~\ref{fig:ebsd-microstructure}d). A Ag precipitate was observed at the interface between the Al$_2$O$_3$ substrate and the Cu film at the GB indicating that Ag diffused along the GB for the entire film thickness.

\subsection{Observations of the GB structure with and without Ag}

\begin{figure}
    \centering
    \includegraphics[width=\linewidth]{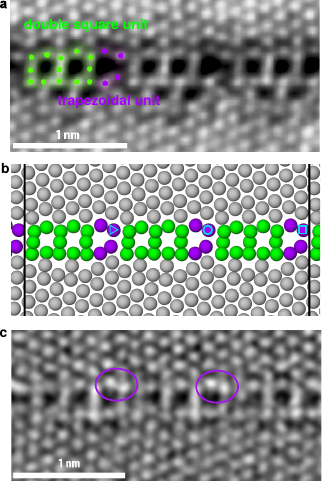}\\
    \caption{The atomic structure of a $\Sigma$ 37c $\langle111\rangle$ $\{3\,4\,7\}$ GB. a) HAADF-STEM image at \SI{300}{kV} of a GB with a misorientation of \ang{50.9\pm0.3} in the pure Cu film. b) MD model of the zipper structure in pure Cu. The vertical lines indicate the unit cell. The triangle, circle and square in petrol show CSL points in different z-heights. c) HAADF-STEM image at \SI{300}{kV} of a GB with a misorientation of \ang{50.9\pm0.3} from the Ag segregated Cu film.}
    \label{fig:zipper-exp-atomic-structure-Cu}
\end{figure}

In a next step, the atomic structure of the grain boundary and the influence of Ag segregation to it were investigated. Fig.~\ref{fig:zipper-exp-atomic-structure-Cu}a shows an atomically-resolved HAADF-STEM image of a $\Sigma$37c $\langle111\rangle$ $\{3\,4\,7\}$ GB in pure Cu with a misorientation of \ang{50.9\pm0.3}. The misorientation angle is an average of at least 10 different measurements of angles between different $\langle220\rangle$ planes of both grains. The bright spots can be directly correlated to atomic columns. The GB structure can be described by the structural unit model~\cite{Han2017} and divided into two alternating sub-units: One double-square unit, consisting of 13 atomic columns, and one trapezoidal unit with 4 atomic columns. The same atomic arrangement was found in the pure Cu film after annealing (see Suppl. Fig.~\ref*{suppl_fig:EBSD-comparison-before_after_annealing}d). Atomistic simulations of the same GB also exhibit the same atomic structure (Fig.~\ref{fig:zipper-exp-atomic-structure-Cu}b).  The total length of the repeating GB structure with both sub-units is equivalent to the length of one third of a $\Sigma$37c $\langle 111 \rangle$ $\{3\,4\,7\}$ CSL lattice cell. The CSL lattice of the investigated GB is shown in Suppl. Fig.~\ref*{fig:DSC-lattice}. After one third, the CSL lattice shows a coincidence site, which is at a different position along the tilt axis direction than the first coincidence site, as indicated by different symbols in the Suppl. Fig.~\ref*{fig:DSC-lattice}. The coincidence lattice sites are as well indicated in petrol in Fig.~\ref{fig:zipper-exp-atomic-structure-Cu}b. In general, a multiple of the length of the repeating GB structure should always be equal to the CSL lattice cell (or an integer multiple of it, in our case it would be 3 times), allowing the repeating GB structure to exist within the repeating CSL lattice. Otherwise, the strain field and coincidence sites would always change and a repeating pattern within the GB would be impossible. The sub-units of the $\Sigma$37c $\langle111\rangle$ $\{3\,4\,7\}$ GB are similar to the $\Sigma$19b $\langle111\rangle$ $\{2\,5\,3\}$ GB~\cite{Meiners2020}, in which the misorientation angle between both grains is \ang{46.8}. In that case, a single square instead of a double-square unit alternates with the same trapezoidal unit as observed in our case. When increasing the misorientation angle between both grains up to \ang{60}, we observed a GB structure which consists only of the square-typed units (see Suppl. Fig.~\ref*{suppl_fig:EBSD-comparison-before_after_annealing}c) similar to a GB with same macroscopic parameters in Aluminum~\cite{Ahmad2023}. Thus, the structural unit model should be valid for this type of symmetric $\langle111\rangle$ tilt GBs with misorientation angles between \ang{46} and \ang{60}.

When investigating a $\Sigma$37c $\langle111\rangle$ $\{3\,4\,7\}$ GB in the Ag annealed Cu film (Fig.~\ref{fig:zipper-exp-atomic-structure-Cu}c), having the same misorientation of \ang{50.9\pm0.3}, the same GB structure can be observed as in the case without Ag: The GB consists of two alternating sub-units, one double-square unit and one trapezoidal. Neither faceting nor phase transitions can be observed for this specific high angle GB upon Ag segregation as it was the case for other GBs in Cu~\cite{Peter2018,Frolov2015}. However, energy dispersive X-ray spectroscopy showed a clear enrichment of Ag at the GB (see Suppl. Fig.~\ref*{fig:Asymmetric_GB}). Meiners et al. investigated Zr segregation at Cu in a similar GB and observed an enrichment of Zr at asymmetric steps~\cite{Meiners2020}. As visible in Suppl. Fig.~\ref*{fig:Asymmetric_GB}, this seems not to be the case for Ag.

\begin{figure}
    \centering
    \includegraphics[width=\linewidth]{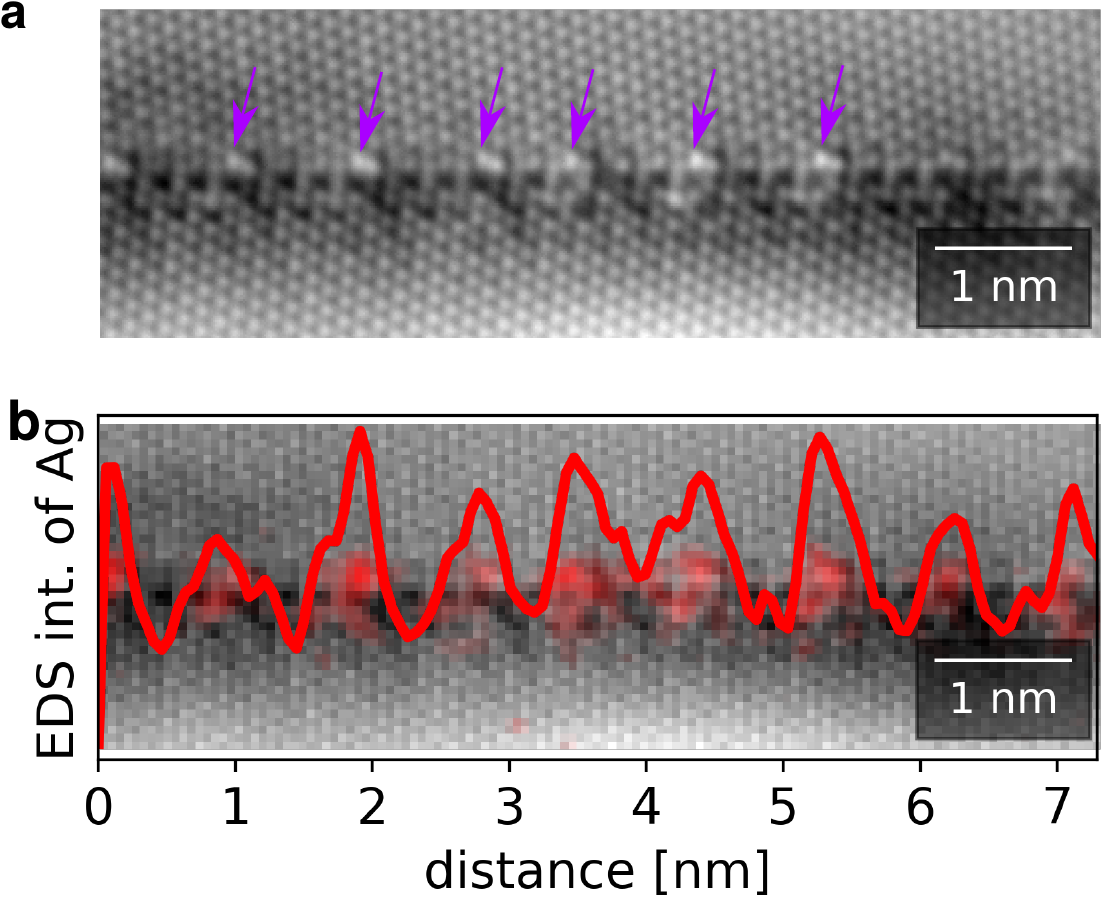}
    \caption{a) HAADF-STEM image at \SI{120}{kV} of a Ag segregated $\Sigma$37c $\langle111\rangle$ $\{3\,4\,7\}$ GB with a misorientation of \ang{50.6\pm0.3}. b) The intensity of the EDS signal of the Ag L-$\alpha$ peak at \SI{2.98}{keV} has maxima at each $\sim$\SI{0.9}{nm} along the GB, coinciding with the trapezoidal units.}
    \label{fig:EDS}
\end{figure}
Having a closer look at the Ag segregated GB, the HAADF-STEM contrast of the atomic columns in the trapezoidal sub-unit seems to be increased (encircled in purple in Fig.~\ref{fig:zipper-exp-atomic-structure-Cu}c). Thus, we decided to investigate the GB with Ag segregation by STEM but using a lower acceleration voltage. At \SI{120}{kV}, the knock-on damage is reduced and less energy is transferred to the Ag atoms, avoiding a beam-induced change of their positions during acquisition and thus blurring of the atomic column contrast~\cite{Egerton2004}. Indeed, a clearer picture can be drawn here (Fig.~\ref{fig:EDS}a): One position within the trapezoidal sub-unit, from now on called "top position 1", has a significantly higher intensity (indicated by purple arrows). As the intensity in a HAADF-STEM image correlates with the atomic number of an element, this implies that Ag -- which has a higher Z number (47) than Cu (29) -- is enriched on that specific position. In a next step, near-atomic resolution energy dispersive X-ray spectroscopy was performed. Fig.~\ref{fig:EDS}b shows the intensity of the Ag-L$\alpha$ peak in the EDS dataset along the GB. Therefore, an integrated Ag line profile was obtained by measuring the Ag-L$\alpha$ peak intensities along the GB with a width of \SI{1.74}{nm} using hyperspy~\cite{hyperspy}. A significant increase in the Ag intensity can be observed at the trapezoidal unit. This supports the HAADF-STEM findings that the trapezoidal units are enriched in Ag.

\subsection{MD/MC simulations and preferential Ag segregation sites}
\begin{figure}
    \centering
    \includegraphics[width=\linewidth]{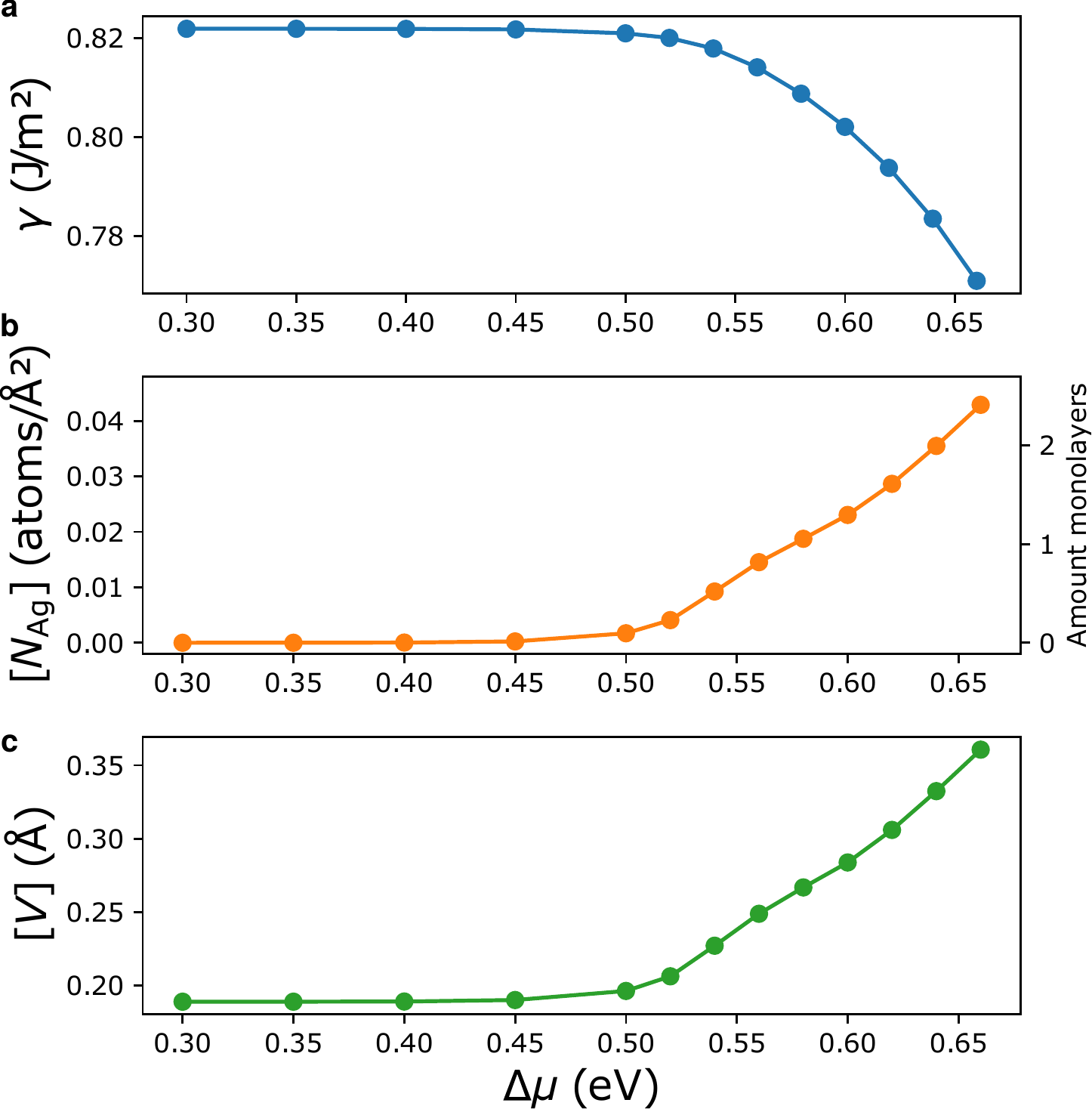}
    \caption{Different thermodynamic properties for increasing chemical potential differences between Cu and Ag. Whereas a) the free GB energy $\gamma$ decreases, b) the Ag excess $[N_\text{Ag}]$ - or equally the amount of monolayers covored by Ag atoms - and c) excess volume $[V]$ of the GB increase with increasing chemical potential difference.} %
    \label{fig:thermodynamic-properties}
\end{figure}
\begin{figure*}
    \centering
    \includegraphics[width=\linewidth]{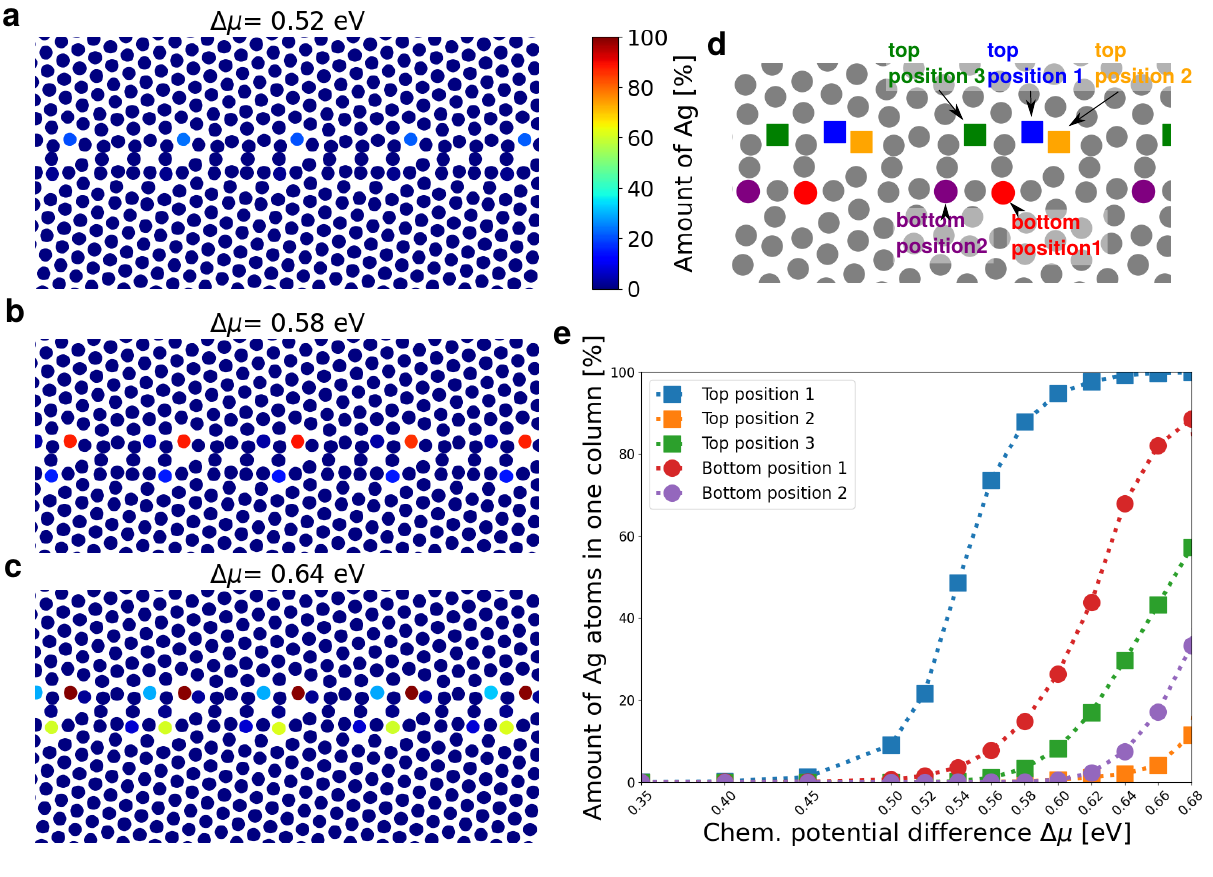}
    \caption{Heat maps of the amount of Ag in each of the atomic columns at 300K are shown for different chemical potential differences -- a) \SI{0.52}{eV}, b) \SI{0.58}{eV}, and c) \SI{0.64}{eV}). d) Schematic showing the five positions within the GB structure to which Ag segregates mostly. e) The fraction of Ag atoms in each column as labelled in d) increases with a higher chemical potential difference.}
    \label{fig:Ag_occupancy}
\end{figure*}

We also performed hybrid MD/MC simulations at \SI{300}{K} with periodic boundary conditions on the same GB. The chemical potential difference between Ag and Cu was increased stepwise, acting as a driving force for Ag to segregate at the GB. In all simulations, as well as in the experiment, we only observed the zipper structure independent of the silver concentration. In order to investigate if GB phase transitions occur, we also performed MD/MC simulations with open boundaries at \SI{900}{K} for \SI{100}{ns} with different chemical potential differences \cite{Frolov2013}. We could once again only observe the zipper structure in the $\Sigma37$c $\langle 111 \rangle$ $\{347\}$ GB.

We calculated the GB free energy as $\gamma = \gamma_0 - \int [N_\text{Ag}] \mathrm{d}\Delta\mu$, with $\gamma_0$ being the GB free energy without Ag excess \cite{Frolov2012a}. The free energy of the GB decreases when the chemical potential difference increases (Fig.~\ref{fig:thermodynamic-properties}a), which signifies a stabilization of the GB upon Ag segregation. In contrast to that, the Ag excess $[N_\text{Ag}]$ of the GB and thus the amount of monolayers occupied by Ag atoms increases (Fig.~\ref{fig:thermodynamic-properties}b). To calculate the monolayers, the Ag excess was divided by the amount of atoms present in one (347) layer (containing only 3 atoms per CSL unit cell due to the high indices of the plane) divided by the area of the layer.
The excess volume $[V]$  increases as well as visible in Fig.~\ref{fig:thermodynamic-properties}c. A major increase in excess volume is the main factor for embrittlement during Bi annealing~\cite{Duscher2004,Keast1998, Schweinfest2004}. In the $\Sigma$5 $[001]$ (310) Cu GB, the excess volume tripled upon Bi segregation~\cite{Lozovoi2008}. In our case, the GB's excess volume only doubled, which may help to explain why Ag segregation does not cause embrittlement~\cite{Lozovoi2006}. It should be noted that other factors, like the local changes in atomic bonding of the GB or kinematic propagation of the crack tip along the GB, also contribute to the embrittlement of an alloy.

As in the experiment, we can observe that Ag substitutes specific GB sites on the atomic level. However, in atomistic simulations we are even able to directly quantify the amount of Ag atoms in each atomic column.
In Fig.~\ref{fig:Ag_occupancy}a-c, the percentage of Ag atoms in each atomic column in the MD/MC simulations is plotted for 3 different chemical potentials. Even when the Ag concentration in the GB increases, the GB structure remains unchanged, with Ag just segregating to new substitutional sites within the structure. At a small chemical potential difference of \SI{0.52}{eV} (Fig.~\ref{fig:Ag_occupancy}a), the top position 1 is the only one which is occupied by Ag atoms with an amount of $\sim$\SI{22}{\%} of Ag out of all atoms in this column. Increasing the chemical potential difference to \SI{0.58}{eV} (Fig.~\ref{fig:Ag_occupancy}b), the top position 1 consists already of \SI{88}{\%} Ag atoms and the bottom position 1 of $\sim$\SI{15}{\%} Ag atoms.
Increasing the difference further to \SI{0.64}{eV} (Fig.~\ref{fig:Ag_occupancy}c), the top position 1 is completely occupied by Ag atoms (\SI{99}{\%}), while the bottom position 1 is occupied by $\sim$\SI{68}{\%} Ag atoms, followed by top position 3 (\SI{30}{\%} Ag atoms), bottom position 2 (\SI{8}{\%} Ag atoms) and top position 2 (\SI{2}{\%} Ag atoms). In Fig.~\ref{fig:Ag_occupancy}e, the amount of Ag at these five specific positions within the GB structure is plotted over the chemical potential difference. These five positions are the five with highest Ag concentrations. First, the top position 1 is getting nearly completely filled, followed by bottom position 1. Then, top position 3, bottom position 1 and bottom position 2 get filled upon a higher chemical potential difference. At chemical potentials higher than \SI{0.68}{eV}, the regular GB structure starts to collapse and Ag precipitates form. We repeated this analysis with datasets equilibrated at elevated temperatures (\SI{500}{K} instead of \SI{300}{K}, see Suppl. Fig.~\ref*{fig:Ag_occupancy_500K}). They follow the same trend at slightly lower chemical potential differences.

The experimental determination of local solute concentrations within a GB structure is to date extremely challenging, but is important for connecting atomistic characteristics of GBs with their thermodynamic ensemble properties. Here, we use the obtained data from atomistic modelling under well defined boundary conditions as a baseline to quantify the atomic scale experimental observations. Two different approaches were used to find the best match for our experiment out of the modelled datasets with varying chemical potential differences. In a first approach, we use the Gibbsian interfacial excess (GIE) obtained from STEM-EDS as an averaged compositional quantity and in a second procedure we quantify the atomic column intensity in the HAADF-STEM images, comparing it to HAADF-STEM image simulations of the modelled datasets with different chemical potential differences. Both methods are explained in detail in the Appendix.

In the first approach, the concentration gradient of Ag is determined by STEM-EDS and averaged across several nanometers (here $\sim$7~nm) of the GB. Then, the concentration of Ag in both grains is subtracted from the concentration gradient, since we are only interested in the Ag excess at the GB. In a next step, the concentration is converted to an atomic density and then integrated to obtain the cumulative GIE, being \SI{1.67\pm0.5}{nm^{-2}}. It can then be compared to the ones from simulations with different amounts of Ag segregation, which were analyzed in the same way. Simulations with chemical potential differences of 0.55 to \SI{0.59}{eV} have a GIE lying in the uncertainty range of the experimental observations. For lower chemical potential differences, less Ag segregates to the GB and thus the GIE would be lower. Is the chemical potential difference higher than \SI{0.59}{eV}, more Ag is present at the GB and thus the GIE is higher compared to the experimentally obtained value. It needs to be mentioned that the method is limited due to channelling effects since the analysis was performed when the sample was in zone axis, which might change concentration values. Also sample thickness and background noise can have an influence on concentration calculations.

In the second approach, the atomic column intensity of the Ag columns at the GB in the HAADF-STEM images is determined and related to the atomistic simulation by STEM image simulations. After subtracting a polynomial background from the atomic resolution image, we normalized the intensities of the Ag columns to the averaged intensity of the Cu atomic columns within a grain to obtain comparable values. The experimental datasets reveal intensities of Ag within the top position 1 varying between 1.06 and 1.42 relative to the intensity of Cu atoms within the grain (see Appendix Fig.~\ref{fig:comparison_HAADF-STEM}c). These values are then compared to HAADF-STEM image simulations of the atomistic models with different chemical potentials. The conditions for the HAADF-STEM image simulation have to be as close as possible to the real experimental conditions, including probe size, collection angles and sample thickness. For chemical potential differences of more than \SI{0.56}{eV}, the intensity of Ag columns compared to Cu columns within the grains, is close to the experimentally observed value. At higher chemical potential differences than \SI{0.58}{eV}, the intensity of the top position 1 does not change significantly since at this point, the column is already completely filled by Ag. However, in higher chemical potential differences, other columns get occupied with Ag and thus the intensities of these columns would increase. Apart from one column (see discussion below), this is not the case in our experiment. Thus, simulations with chemical potential differences higher than \SI{0.6}{eV} would not match to our experiment.

To conclude, both methods which can compare simulation and experiment reveal that simulations with chemical potential differences between \SI{0.55} to \SI{0.59}{eV} would fit best to our experimental observations.

In Fig.~\ref{fig:EDS}a, one can also observe that different columns of the top position 1 have different HAADF-intensities. These values are normalized intensities relative to the average intensity of atomic columns within the grains. Knock-on damage, which can promote the re-distribution of atoms and thus leading to variations in intensities during longer acquisition times have been minimized by acquiring the data at \SI{120}{kV}. Other factors, such as contamination, build-up or sputtering of atoms at the GB can also lead to errors in the calculation of atomic column intensities. However, the intensity changes observed varying on a column-by-column basis over a length of about \SI{7}{nm} can be as well related to local changes in Ag concentration. Furthermore, one bottom position 1 is visibly brighter. It is indicated in Appendix Fig.~\ref{fig:comparison_HAADF-STEM}c in the center of the image and has a value of 1.24. This value is substantially higher than the values of neighboring bottom positions 1, which have an intensity of 0.96, which correlates to a brightness intensity of a Ag-free column. This indicates that only at that specific position, the Ag intensity might be as well increased, indicating that the bottom position 1 is also partially filled by Ag. Furthermore, this observation underlines that local intensity changes might be present.

In the next section, we want to elucidate the reason for Ag occupying these specific positions.

\subsection{Influence of GB structure on preferential segregation sites}

\begin{figure*}
    \centering%
    \includegraphics{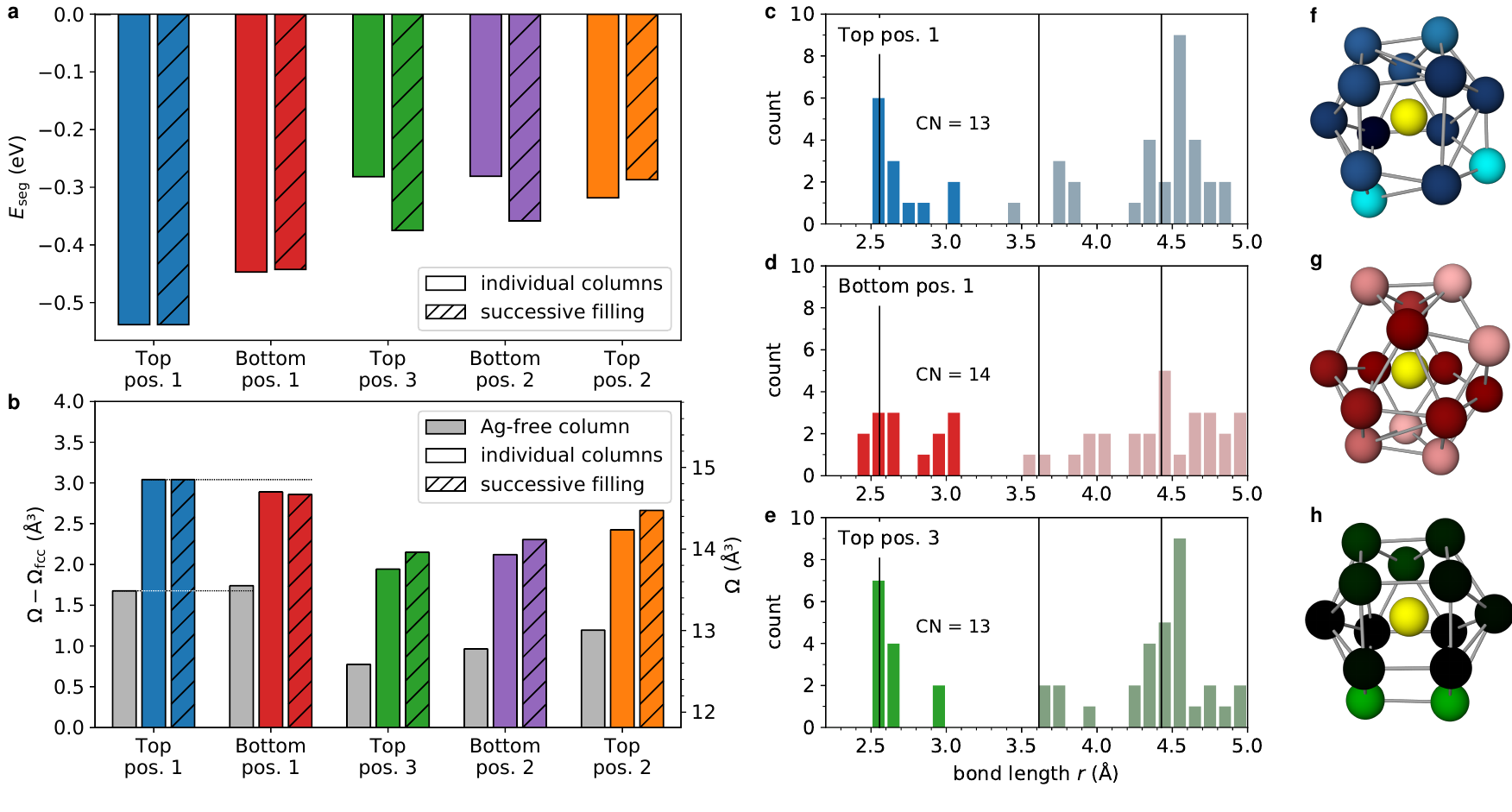}%
    \caption{Properties of different GB sites. a) The segregation energy of each position. The solid bars represent individual segregation energies and the hatched bars represent segregation energies when successively filling the positions (e.g., the hatched green bar shows the segregation energy when top position 1 and bottom position 1 are already filled).   b) Local Voronoi excess volumes (atomic volume $\Omega$ for each column minus the atomic volume of fcc atoms $\Omega_\text{fcc}$). The absolute volume is shown on the right axis. For the Ag-free system (gray bars),  bottom position 1 has a slightly higher volume compared to top position 1 ($\Delta\Omega = \SI{0.06}{\angstrom^3}$). Top position 1, however, expands to a higher atomic volume ($\Delta\Omega = \SI{-0.15}{\angstrom^3}$) upon insertion of Ag than bottom position 1 (colored bars). c-e) Diagram showing the radial distances and amount of next atoms for the first three positions getting occupied. The radial distances are measured in pure Cu GBs. The black lines indicate the neighbor distances in defect-free fcc. Darker bars show bond shorter than \SI{3.25}{\angstrom}, which corresponds roughly to the first neighbor shell. The corresponding coordination numbers (CN) are also shown.  f--h) First neighbor shell of an atom (yellow) in top position 1 (f), bottom position 1 (g) and top position 3 (h). The brightness indicates the bond length of the neighbor atom (dark being \SI{2.55}{\angstrom} and light being \SI{3.10}{\angstrom}).}
    \label{fig:position-properties}
\end{figure*}
In the simulation, we observed a step-by-step filling of different atomic sites in the GB when increasing the chemical potential difference. We could correlate two of the states with our experimental observations, in which one atomic site (called top position 1) is filled nearly completely and a second atomic site (bottom position 1) is only partially filled. In the experiment, we can further observe site-by-site intensity variations in the HAADF intensity, which might be correlated to local changes in Ag occupancy of each atomic column. 
To understand the preferential site occupation of Ag atoms in the GB structure, we investigated different parameters such as the atomic volumes, distances to nearest neighbor atoms, Steinhardt parameter~\cite{Steinhardt1983}, and segregation energies of the different positions.

To this end, we first calculated the segregation energy of each position, defined as the energy difference of a system with Ag atoms on a GB site compared to a system with Ag atoms in the bulk (Fig.~\ref{fig:position-properties}a)~\cite{Vitek1982}. Therefore, we replaced the complete column of each segregation site in the GB individually with Ag (solid bars in Fig.~\ref{fig:position-properties}a) and we could observe a negative segregation energy for all positions. We also tested segregation energies for implanting isolated Ag atoms to the investigated positions, but only small segregation energy differences compared to a completely filled column could be observed (Supplemental Figs.~\ref*{fig:Eseg-atom-vs-col}~and~\ref*{fig:sideview}).  As expected, the top position 1, which gets occupied by Ag atoms first in the simulation and is also occupied by Ag in the experiment, is the one with the lowest segregation energy, followed by the bottom position 1. Interestingly, it seems that the top position 2 should be the third column being filled up based on the segregation energies, but it is the fifth and last column where Ag segregates to. The hatched bars in Fig.~\ref{fig:position-properties}(a) show the segregation energies for successive filling of the columns as it was observed in the simulation, i.e., the segregation energy of bottom position 1 is calculated for a sample system with filled top position 1 and so on. For top position 2, the segregation energy for Ag atoms to this position was calculated in a system in which the other four columns were already occupied. Under these circumstances, the segregation energies are consistent with the results in Fig.~\ref{fig:Ag_occupancy}. 

As a different potential parameter for predicting preferential segregation sites, we investigated the free volume of each position in the GB structure using a Voronoi analysis~\cite{Voronoi1908}. 
Findings for the $\Sigma$5 $\langle111\rangle$ GB in Cu~\cite{Duscher2004,Huang2020} or $\langle110\rangle$ GBs in Ni with Pd segregation~\cite{Rittner1997} suggest that for bigger solute elements, the position in the GB structure with the highest volume would be the preferential segregation site. With the EAM potential, we found an excess Voronoi volume of \SI{0.99}{\angstrom^3} for a substitutional Ag point defect, indicating that Ag atoms are clearly larger than Cu atoms and might prefer segregation sites with high free volume. In the investigated GB structure, the positions with the highest atomic volume in a pure Cu structure are top position 1 with \SI{13.72}{\angstrom^3} and bottom position 1 with \SI{13.79}{\angstrom^3}, followed by top position 2 with \SI{13.21}{\angstrom^3} (Fig.~\ref{fig:position-properties}b, gray bars). The bigger atomic volume of top and bottom position 1 compared to the rest of the sites can explain the preferential segregation to these two sites. Even though their atomic volumes are similar to each other, the top position 1 is occupied by Ag atoms prior to bottom position 1. 

Differences in atomic strain at different sites can also be a factor for preferential site occupation~\cite{Nie2013,Kaeshammer2015}. Thus, we investigated the volume expansion of all five segregation sites upon Ag filling of each site individually (filled colored bars in Fig.~\ref{fig:position-properties}b) or upon successive filling (hatched bars) following the order shown in Fig.~\ref{fig:Ag_occupancy}e. We can observe that top position 1 has a slightly larger volume than bottom position 1 after being filled with Ag, even though there was a small tendency in the opposite direction in the pure Cu case (bottom position 1 was slightly larger compared to top position 1). This is true regardless whether top position 1 was already filled by a Ag atom (hatched bar) or only bottom position 1 (colored bar) is occupied by Ag. It indicates that top position 1 can expand more easily compared to bottom position 1, which can lead to a preferential position. However, looking at the atomic volumes of the consecutive positions, the expected trend of decreasing atomic volumes for less favorable positions cannot be observed. According to the atomic volumes, top position 2 should be the position being occupied after top and bottom position 1. However, it is top position 3, with the lowest atomic volume both in a pure Cu cell and when it is occupied by Ag. Thus, the atomic volume alone is insufficient to explain the preferential segregation sites.

In a next step, we also had a look at the specific atomic environments of different positions.
Therefore, we investigated the bond length of all neighboring atoms as shown in Fig.~\ref{fig:position-properties}c--h for the first three positions. In plots c--e it is shown that the closest atoms are present in bond distances between \SI{2.4/2.5}{\angstrom} to \SI{3.1}{\angstrom}. In none of the three investigated positions, atoms occur with bond distances of 3.1 to \SI{3.4}{\angstrom}, but only again at higher distances than \SI{3.4}{\angstrom}. In a bulk Cu lattice, the nearest neighbor atoms have a distance of \SI{2.55}{\angstrom} (indicated by a black line). In bottom position 1, two atoms are even slightly closer than that. This might be unfavorable for Ag due to its atomic size mismatch, but due to the delocalized nature of the metallic bond it is difficult to draw definite conclusions from the first neighbor shell environment alone. All in all, no striking, qualitative differences between the three diagrams are visible that could serve as a simple predictor for preferential segregation sites.

Furthermore, Scheiber et al. found a linear correlation between the 5th Steinhardt order parameter \cite{Steinhardt1983} and the segregation energy~\cite{Scheiber2015}. We therefore also tried this analysis (Supplemental Fig.~\ref*{fig:steinhardt}, calculated using the software pyscal \cite{Menon2019}), but could find no such correlation for any of the first 12 parameters. This suggests that the changes in the bonding environment are relatively delocalized in our case so that they are not easily detectable with methods that consider only the next-neighbor environment. Therefore, the bond lengths and number of nearest neighbor atoms or the shape of the polyhedron are not suitable as simple predictors to determine preferred segregation sites within a GB in a metallic system in accordance with the observations of Ziebarth et al. for Fe impurities in Si GBs~\cite{Ziebarth2015}.  %

Furthermore, preferential sites of solutes were also identified due to dangling bonds or changes of covalent bonds especially in ceramic materials or semimetals~\cite{Zhao2019,Buban2006,Fabris2003,Chen2005}.
Upon changes in the electronic structure, the cohesive energy also changes and can thus lead to embrittlement. Pronounced effects can be observed if the solute atoms differ in electronegativity compared to the host element~\cite{Briant1990,Messmer1982}. However, for Ag in Cu, both elements have similar electronegativity and embrittlement of Cu due to Ag segregation has not been observed. Since both are metals, no dangling or covalent bonds are expected and changes in the electronic band structure are not accessible with the methods used in the present work.%

We can conclude that besides the segregation energy, no simple predictor of the preferential segregation sites could be found. While local free volume plays a role, local bonding effects cannot be neglected and depend on the complex atomic environment of the GB. However, these are not easy to compute and thus not suitable for simple predictions.

\section{Conclusion}
We investigated the segregation behavior of Ag in a symmetric tilt GB in Cu and different parameters which might affect the site occupation combining experimental observations and atomistic simulations.
The atomic structure of a $\Sigma$37c $\langle111\rangle$ $\{3\,4\,7\}$ GB in copper was resolved using HAADF-STEM. The same structure was replicated in MD simulations. After Ag alloying, Ag atoms only substitute specific positions in the GB. Neither faceting nor phase transition was observed. %
Comparing experiment and hybrid MD/MC simulations using HAADF-STEM intensity changes and Gibbsian interfacial excess determination, we conclude that simulations with a Ag excess of \SI{0.015}{atoms\angstrom^{-2}} to \SI{0.019}{atoms\angstrom^{-2}} match best to our experiment (equivalent to a chemical potential difference of $\Delta \mu =$ \SI{0.56}{meV} to \SI{0.58}{meV}). In these cases, two columns in the GB structure are occupied with Ag, whereas top position 1 is mostly filled and bottom position 1 only partially. In the experiment, local changes of Ag occupation between columns are observed, which can induce site-specific changes in GB properties. %
We investigated the relevance of different parameters (volume, configuration of surrounding atoms, Steinhardt order parameter) for predicting preferential segregation sites in the GB structure. 
The preferential segregation sites seem not be solely determined by their local volume of each site and their expansion upon segregation. Also the bond lengths of surrounding atoms and the Steinhardt order parameter do not show a simple clear trend following the order in which the positions are filled up according to simulations. Only the calculation of the segregation energy, which takes into account the local atomic bonding in each position, follows the same trend as the occupation of Ag in specific sites.
A combination of both experiment and simulation as presented above is needed to determine the atomic level distribution of solutes at GBs and for understanding the atomistic origin of GB segregation. These two ingredients ultimately provide the link to the thermodynamic characteristics of the interface and how they affect macroscopic material properties.

\section{Data availability}
The main datasets of this study are published under \url{https://doi.org/10.5281/zenodo.7372733}.

\section{Acknowledgements}

We gratefully acknowledge G.~Richter's team from the Max
Planck Institute for Intelligent Systems for producing the Cu thin film and T.~Oellers for the Ag deposition at the Ruhr-Universit\"at Bochum. The authors thank N.~Peter for helpful discussions and initiating this project by supporting in Ag deposition on the film.  This project has received funding from the European Research Council (ERC)
under the European Union's Horizon 2020 research and innovation
programme (Grant agreement No.~787446; GB-CORRELATE). 

Author contributions: T.B. and L.L. contributed equally to this work. L.L. performed the experimental sample preparation, HAADF-STEM investigations and analysis of the obtained datasets. C.H.L. and G.D. designed concept of the experimental study.  The simulations and analyses of the simulation data were performed by T.B.. G.R. provided the Cu thin film, the main material used in this study. The project was supervised by C.H.L. and G.D., who also contributed to discussions. G.D. secured funding for L.L. and T.B. via the ERC grant GB-CORRELATE. L.L. and T.B. prepared the initial draft and all authors contributed to the preparation of the final manuscript.

\appendix
\section{Comparison of experiment and simulation to quantify the Ag excess}

We used two methods to determine the local Ag excess at the GB by comparing experiment with atomistic simulations. In the first method, we determine the Gibbsian Interfacial Excess (GIE) from the STEM-EDS and compared it to the GIE obtained from MD/MC simulations. In the second method, we measure the atomic column intensity in HAADF-STEM images of the top position 1 in the GB relative to the bulk Cu columns and compared it to HAADF-STEM image simulations using the MD/MC cells as input.

\subsection*{Method 1: Determining the Gibbsian interfacial excess from STEM-EDS}

\begin{figure}
    \includegraphics[width=\linewidth]{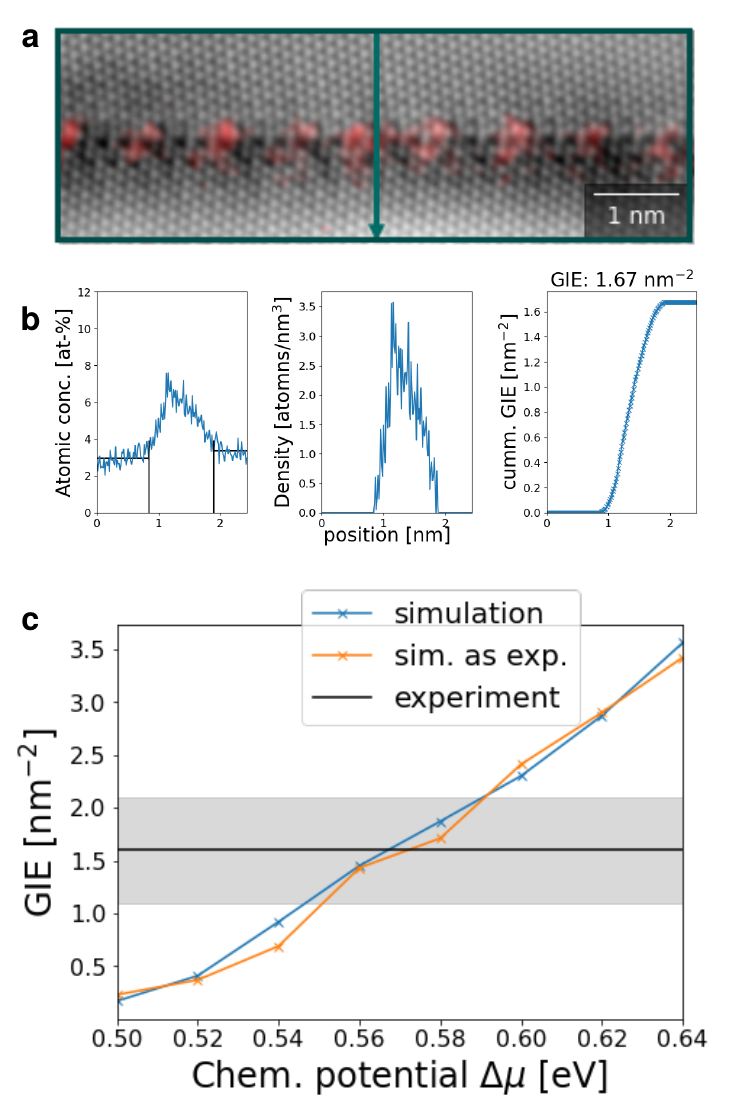}
    \caption{Comparing the calculated gibbsian interfacial excess (GI) value of simulations with the experiment. a) HAADF-STEM image with EDS signal of Ag overlaid. The concentration profile of Ag was measured along the arrow, with a width equal to the frame size. b) Diagrams showing the steps needed to convert the concentration to the GIE. c) Comparison of the GIE of simulations with experiments. All simulation datasets for the different chemical potentials were equilibrated and performed at \SI{300}{K}.}
    \label{fig:comparison_GIE}
\end{figure}

We determine the Gibbsian interfacial excess (GIE) from STEM-EDS datasets similar to the procedure described in Refs.~\cite{Krakauer1993,Maugis2016}. We used the EDS maps acquired at \SI{120}{kV} to reduce knock-on damage and to maximize the X-ray yield. In a first step, the Ag content was quantified using the Velox Software (vers. 3.5). For the quantification, a standard-less Cliff-Lorimer quantification with absorption correction was used, which takes into account the actual geometry of the detectors and the stage tilt. The sample thickness was set to \SI{50}{nm} and a Brown-Powell ionization cross-section model was used.  In a second step, the Ag concentration profile across a GB with a width of \SI{7}{nm} parallel to the GB plane -- equivalent to about 7 structural units of the $\Sigma$ 37c $\langle111\rangle \{347\}$ GB -- was summed. Such a concentration profile is shown in Fig.~\ref{fig:comparison_GIE}b on the left side. Since we are only interested in the Ag excess at the GB, thus the surplus of Ag content, we subtracted the Ag concentration in the grains. In order to access the GIE, the Ag concentration needs to be converted to an atomic density. Therefore, the density of fcc Cu along the $\langle 111 \rangle$ direction was used, being \SI{84.67}{atoms nm^{-2}} for a lattice parameter of $a = \SI{3.615}{\angstrom}$. This is shown in Fig.~\ref{fig:comparison_GIE}b in the middle. The integral of the atomic density is then equal to the GIE (Fig.~\ref{fig:comparison_GIE}b right side). An experimental value of the GIE of about \SI{1.67\pm0.5}{nm^{-2}} was experimentally determined for 2 independent EDS data sets acquired at different GB positions.

For estimating the errors present in our calculation, the averaged Cu density as well as the integration width were changed. Thereby, changing the atomic density by 6$\%$ can lead to an uncertainty in the GIE of up to 22$\%$. A decrease of 30$\%$ of the integration width can lead to a decrease in GIE of 15 $\%$. Furthermore, the quantification of Ag concentration by EDS has an uncertainty of $\pm 2\,\text{at-\%}$. To conclude, the method presented above is sensitive to many parameters affecting its accuracy to determine the GIE. Thus, we used the experimentally measured GIE only as a rough guideline to evaluate which range of simulated Ag excess values and hence chemical potential difference match to the experiment. Therefore, we performed the same analysis as described above for the simulated datasets, even though the GIE could be extracted from the simulated datasets in a direct manner as shown in Fig.~\ref{fig:thermodynamic-properties}b. We compared the directly obtained GIE for Ag in the simulated datasets with the ones calculated analogous to the experiment and observed deviations of max. 5$\%$. This shows that the pathway of obtaining the GIE is fairly correct and only experimental uncertainties (like the absolute quantification of Ag and defining the integration width of the Ag peak) can affect the experimentally extracted GIE. 
As expected, with increase of a chemical potential difference, the GIE increases since more Ag diffuses into the system and also more Ag is present at the GB. The simulations overlap with the experiments for chemical potential differences between \SI{0.55}{eV} to \SI{0.59}{eV}.

\subsection*{Method 2: Quantifying the intensity of Ag columns from atomic resolution using the HAADF-STEM images}

\begin{figure}
    \centering
    \includegraphics[width=\linewidth]{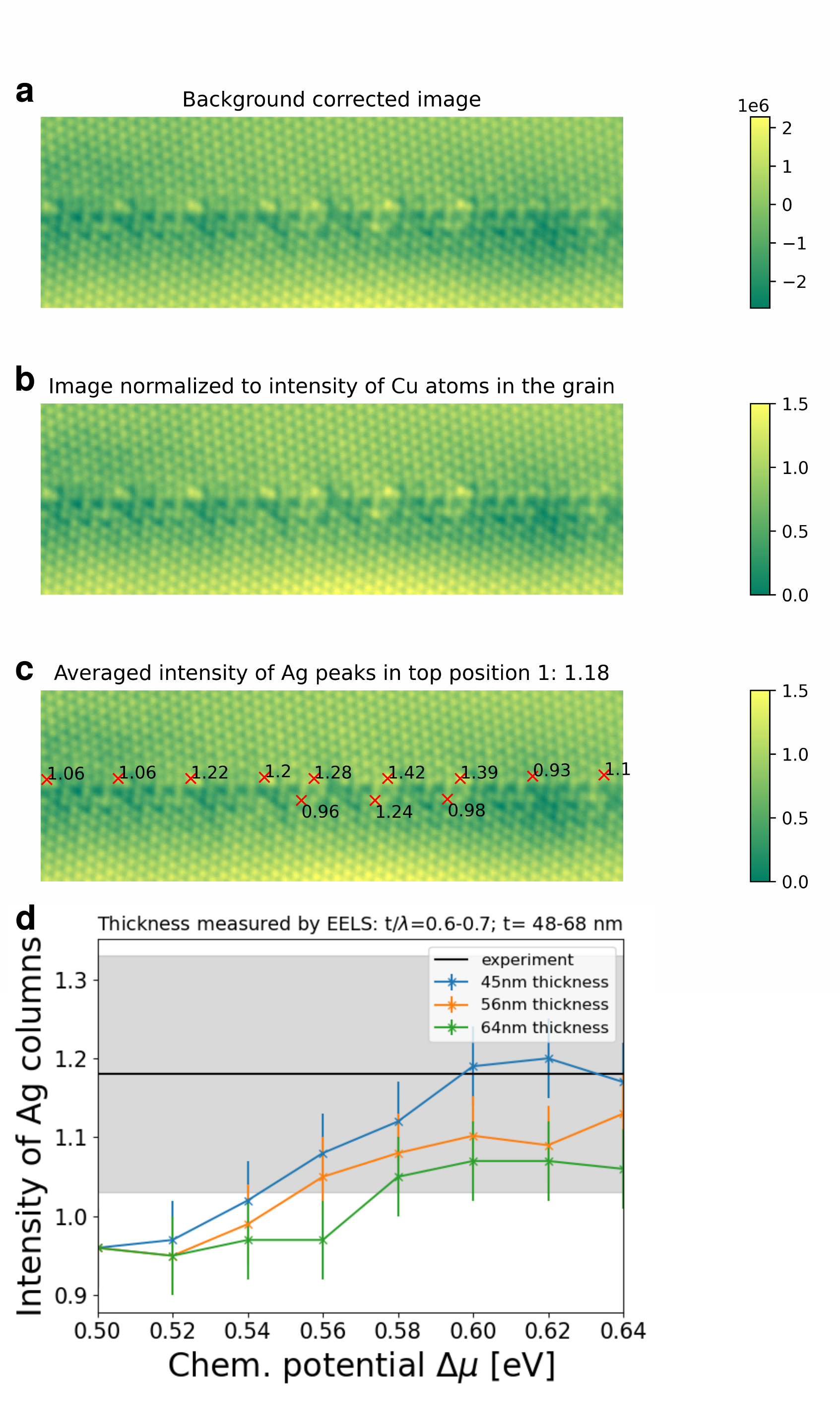}
    \caption{Comparing the HAADF-STEM intensities of atomic columns enriched with Ag with the HAADF-STEM intensity of Cu columns in the grain. a) Intensity of a background corrected HAADF-STEM image. b) The intensity values of all columns were measured and normalized by the average intensity values of atomic columns in the grains. c) Intensity values of the Ag enriched columns. d) A comparison was performed for the experimental image with HAADF-STEM image simulations, using the same imaging conditions as used in the experiment. A thickness range for the simulation cells was used to take into account uncertainties for the experimentally determined thickness value of the sample.}
    \label{fig:comparison_HAADF-STEM}
\end{figure}

In order to estimate the amount of Ag in atomic columns within the GB, we quantify the atomic column contrast in the HAADF-STEM images. Due to the complex atomic arrangements of atoms at the GB core, we determine the intensity of the Ag-rich columns relative to atomic columns within the grain as a reference. The experimental datasets are then compared to HAADF-STEM image simulations using the MC/MD simulation cells as input. In a first step, a 2D-polynomial background subtraction was performed on the experimental images (Fig.~\ref{fig:comparison_HAADF-STEM}a). Then, a peak finding algorithm of the python package ATOMAP (vers. 0.3.1) was used to determine the position of each of the atomic columns and their single peak intensity values~\cite{Nord2017}. The intensity values got normalized by the average of the intensity of atomic columns located in both grains (see Fig.~\ref{fig:comparison_HAADF-STEM}b). In a final step, the intensity of the top position 1 within the GB structure was extracted. The intensities in the experimental image vary between 1.06 and 1.39, excluding the minimum and maximum since they might stem from non-accurate background substraction. Thus, in average, the top position 1 has an intensity of $1.18 \pm 0.15$ as shown in Fig.~\ref{fig:comparison_HAADF-STEM}c. Besides of one bottom position 1, which has an intensity value of 1.24, the intensity of all other positions does not show a substantial increase from the averaged value of 1 and are thus not listed here. This method also exhibits several uncertainties, as the intensity value might be influenced in the first place by the sample thickness, amorphous layers on the surface of the sample or oxides and background noise, and in the second step by using different background correction methods or extracting the intensity of the atomic columns with other algorithms. Thus, great care has to be taken and the final results can be only understood as a rough estimation.

For the comparison with MD/MC simulations, a simulation cell of a similar thickness as the lamella in the experiment is needed. We estimated the thickness of the investigated sample to be between 55 and \SI{64}{nm} using EELS and the thickness determination according to Malis et al.~\cite{Malis1988}. In a next step, we took the MC/MD cells and replicated them periodically along the $\langle111\rangle$ direction to the desired thickness. In a next step, we used these cells to create synthetic HAADF-STEM images using the multislice algorithm with parameters which match the experimental parameters as good as possible (an aberration-free electron probe with \SI{120}{kV}, semi-angle of \SI{24}{mrad}, focal spread of \SI{60}{\angstrom}, an annular detector detecting  with a collection angle range of  95 to \SI{200}{mrad}, corresponding to the setting of our HAADF detector in the experiment, and a beam scan step size of \SI{0.14}{\angstrom}, which corresponds to the pixel size). The image simulation was performed with the python library abTEM vers. 1.0.0b11~\cite{abtem}.

Having the simulated HAADF-STEM images for each chemical potential, we followed the same steps as for the experimental images. Only the first step of background correction was not necessary since the contribution from background is negligible in simulation. All atomic column positions and their peak intensities were determined using ATOMAP. The intensity of all columns was normalized by the average intensity of the atomic columns within the bulk grains. In Fig.~\ref{fig:comparison_HAADF-STEM}d, the change in intensity ratio of the top position 1 in dependence of chemical potential difference is shown for different simulation cell thicknesses. The average intensity ratio determined from the experiment is given as black line and the gray shaded area represents the uncertainty range considering different columns. The intensity values of top position 1 from the simulated images start to overlap with the ones from experiments for chemical potential differences at about \SI{0.56}{eV} to \SI{0.58}{eV}, depending on the simulated thickness. When increasing the chemical potential difference further, we can see that other positions (as bottom position 1, 2 and top position 2,3), get occupied by Ag and thus increase in intensity values. In the experiment, the HAADF STEM intensities of top position 2,3 and bottom position 1,2 do not show significantly higher values than other positions within the GB (besides bottom posiiton 1 in one position as indicated in Fig.~\ref{fig:comparison_HAADF-STEM}c). For simulated datasets with a chemical potential difference of more than \SI{0.60}{eV}, a significant increase in HAADF STEM intensity in other atomic columns is detected. Thus, when comparing HAADF-STEM intensities of simulations and experiment, simulations using a chemical potential difference of \SI{0.56}{eV} and \SI{0.60}{eV} are matching best with the experiments.

To conclude, both independent methods predict that a chemical potential difference between \SI{0.56}{eV} and \SI{0.60}{eV} would fit best to our experimental dataset.

\vfill

\end{document}